\documentclass[a4paper,10pt]{article}

\usepackage[round,authoryear,sort]{natbib}
\usepackage{graphicx}
\bibpunct{(}{)}{,}{a}{,}{,~}
\newcommand{\forget}[1]{}
\usepackage{amsmath}

\newcommand{\ket}[2]{|#1\rangle _{#2}}

\newcommand{\up}{\!\uparrow}
\newcommand{\down}{\!\downarrow}

\title{Bell on Bell's theorem: The changing face of nonlocality} 

\author{ Harvey R Brown\thanks{harvey.brown@philosophy.ox.ac.uk} \\ \textit{Faculty of Philosophy and Wolfson College, Oxford}\\ Christopher G Timpson\thanks{christopher.timpson@bnc.ox.ac.uk} \\ \textit{Faculty of Philosophy and Brasenose College, Oxford}}

\begin{document} 
\maketitle

\begin{abstract}
\noindent Between 1964 and 1990, the notion of nonlocality in Bell's papers underwent a profound change as his nonlocality theorem gradually became detached from quantum mechanics, and referred to wider probabilistic theories involving correlations between separated beables. The proposition that standard quantum mechanics is itself nonlocal (more precisely, that it violates `local causality') became divorced from the Bell theorem \textit{per se} from 1976 on, although this important point is widely overlooked in the literature. In 1990, the year of his death, Bell would express serious misgivings about the mathematical form of the local causality condition, and leave ill-defined the issue of the consistency between special relativity and violation of the Bell-type inequality. In our view, the significance of the Bell theorem, both in its deterministic and stochastic forms, can only be fully understood by taking into account the fact that a fully Lorentz-covariant version of quantum theory, free of action-at-a-distance, can be articulated in the Everett interpretation.

\end{abstract}

\section{Introduction}

John S. Bell's last word on his celebrated nonlocality theorem and its interpretation appeared in his \citeyear{bell:nouvelle} paper `La nouvelle cuisine', first published in the year of his untimely death. Bell was careful here to distinguish between the issue of `no-superluminal-signalling' in quantum theory (both quantum field theory and quantum mechanics) and a principle he first introduced explicitly in 1976 and called `local causality' \citep{bell:tlb}. In relation to the former, Bell expressed concerns that amplify doubts he had already expressed in 1976. These concerns touch on what is now widely known as the \textit{no-signalling theorem} in quantum mechanics, and ultimately have to do with Bell's distaste for what he saw as an anthropocentric element in orthodox quantum thinking. In relation to local causality, Bell emphasised that his famous factorizability (no-correlations) condition is not to be seen `as the \textit{formulation} of local causality, but as a consequence thereof', and stressed how difficult he found it to articulate this consequence.  He left the question of any strict inconsistency between violation of factorizability and special relativity theory unresolved, a not insignificant shift from his thinking up to the early 1980s. Bell felt strongly that correlations ought always to be apt for causal explanations---a view commonly attributed to the philosopher Hans Reichenbach (though perhaps in this particular context incorrectly---of which more below). But he recommended that factorizability `be viewed with the utmost suspicion'.

Few if any commentators have remarked on this late ambivalence about the factorizability condition on Bell's part. Part of the problem facing him was, as he himself emphasised, that although the intuition behind local causality was based on the notion of cause, factorizability makes no reference to causes. Whatever else was bothering Bell, apart from the absence of a clear-cut relativistic motivation, is perhaps a matter of speculation. We suggest, at any rate, that there is an important difference between the notion of locality he introduced in his original 1964 theorem and that of local causality. The difference is essentially that in 1964 he was excluding a certain kind of instantaneous action-at-a-distance, whereas the connection between local causality and such exclusion is not straightforward. More bluntly, and counter-intuitively: violation of local casuality does not necessarily imply action-at-a-distance. The issue is model-dependent. There is a consistent Lorentz covariant model of quantum phenomena which violates local causality but is local in Bell's 1964 sense: the Everett picture. One of the themes of our paper is something that Bell himself emphasised, namely, the consideration of detailed physical models is especially important in correcting wayward intuitions.

The importance of the distinction between Bell's 1964 and 1976 versions of his nonlocality theorem has been noted in the literature, particularly in the recent careful work of Wiseman (\citeyear{Wiseman:2014}; see also \citet{timpsonandbrown:2002}). But in the present paper we are more concerned with the notion of nonlocality than with the theorems \textit{per se}; appreciation of the changing face of nonlocality in Bell's writings, and its wider significance in the foundations of quantum mechanics, is harder to find in the literature. In fact, there are categorical claims that Bell's understanding of nonlocality never underwent significant revision.\footnote{See e.g. \citet{maudlin:what}, \citet{Norsen:2011}, and \citet{Wiseman:2014} for further discussion.} We think this unlikely, but more importantly, we wish to concentrate on how the 1964 and post-1976 notions of nonlocality \textit{should} be understood. 

The change in Bell's definition of locality was accompanied by a shift of thinking on his part and that of many other commentators from the 1970s onwards, in relation to the range of applicability of the Bell theorem. Originally confined to deterministic hidden variable interpretations of quantum mechanics, the theorem was later correctly seen to apply to probabilistic theories generally. Thus it is a widely held view in the literature today that as a result of the predicted (and experimentally corroborated) violation of Bell-type inequalities, standard quantum mechanics itself is nonlocal.\footnote{Consider the following statements in the recent physics literature that arguably are representative of a widespread view concerning the significance of the Bell nonlocality theorem. The first appears in a lengthy 2013 paper by A. Hobson on the nature of quantum reality in the \textit{American Journal of Physics}:
`Violation of Bell's inequality shows that the [entangled state]...is, indeed, nonlocal in a way that cannot be interpreted classically.' \citep[p.220]{Hobson:2013a}
In his reply to critics, Hobson writes: `This violation [of the Bell inequality] means that the correlations are too tightly dependent on the non-local phase relationship between the two [entangled] systems to be explainable by purely local means. So such correlations do lead to nonlocality as a characteristic quantum phenomenon.' \citep[pp.710--711]{Hobson:2013b}.

The final quote is the opening sentence in an extensive 2014 multi-authored paper on Bell nonlocality in \textit{Reviews of Modern Physics}:
`In 1964, Bell proved that the predictions of quantum theory are incompatible with those of any physical theory satisfying a natural notion of locality.'  \citep[p.420]{Brunneretal:2014} 
In fact, Bell did no such thing in 1964; his original version of the theorem refers only to deterministic hidden variable theories of quantum mechanics (see below). The definition of locality that the authors of the 2014 review paper actually give reflects that found in Bell's later papers, where determinism is no longer involved.}
And yet, as Bell himself acknowledged, the proof that standard quantum mechanics violates local causality need not rest on the Bell theorem---or even entanglement. A proof is possible which is closely related to Einstein's pre-1935 argument for nonlocality in quantum mechanics.

We will attempt to spell these points out in more detail below by analysing the development of Bell's remarkable work from 1964 to 1990, and discussing the relevance of the modern Everettian stance on nonlocality.

\section{The original 1964 theorem: Lessons of the de Broglie--{B}ohm theory}

The original version of Bell's nonlocality theorem was the fruit of a penetrating review of deterministic hidden variable theories published in \citeyear{bell:1966}, regrettably several years after it had been submitted for publication.\footnote{For further details see \citet{Wiseman:2014}.} Bell achieved many things in this paper, not all of which have been duly recognised. In particular, there is still insufficient appreciation in the literature that in this paper, Bell was the first to prove the impossibility of non-contextual hidden variable theories, and that his interpretation of this result was the polar opposite to that of Simon B. Kochen and Ernst Specker, who independently and famously published a finitist, but much more complicated version of the proof in \citeyear{kochenspecker}. As Bell saw it, what he had shown was that a fragment of the contextualism built into the de Broglie--Bohm (pilot wave) hidden variable theory has to be found in any deterministic hidden variable theory consistent with the (at least state-independent) predictions of quantum mechanics, and that this is all well and good. In contrast, Kochen and Specker thought they had actually ruled out all hidden variable theories!\footnote{See \citet{brown:1992} and \citet{mermin:1993}.} Another interesting feature of Bell's thinking is this. Although he did not use the result in his simple version of the proof of the Bell-Kochen-Specker theorem, Bell realised that Gleason's  theorem \citep{gleason:1957}, and in particular its corollary concerning the continuity of frame functions on Hilbert spaces, ruled out non-contextualist hidden variable theories for most quantum systems. What Bell did not suspect in 1964, even so, was that the Gleason theorem could also be used to derive a nonlocality theorem for correlated spin-1 systems involving no inequalities.\footnote{See \citet{heywoodandredhead:1983}, \citet{stairs:1983}, \citet{brownandsvetlichny:1990} and \citet{elby:1990}. The Heywood-Redhead and Stairs results predate Greenberger-Horne-Zeilinger \citeyearpar{GHZ} in terms of providing a Bell theorem with no inequalities, though this is widely overlooked in the literature; for example \citet{blaylock} footnote 35, and \citet{Brunneretal:2014} make no reference to these early spin-1 theorems. In the case of bipartite systems, the tight connection between contextualism and nonlocality means that the Bell-Kochen-Specker theorem raises a \textit{prima facie} consistency problem for the very assumptions in the original 1964 Bell theorem as applied to quantum mechanics. To see how nonetheless consistency is assured, see \citet{brown:1991}.

Nor did Bell suspect in 1964 that a weaker (state-dependent) version of the Bell-Kochen-Specker theorem for a single quantum system could be given in which a Bell-type inequality is shown to be both a consequence of non-contextualism and violated by quantum mechanics. This result is originally due to \citet{homeandsengupta:1984}. They considered the case of a single valence electron, say in an alkali atom, in the $^2 P_{\frac{1}{2}}$ state, in which the electron wavefunction involves entanglement between spin and orbital angular momentum. A more manifestly self-consistent version of the proof was given in \citet{fosterandbrown:1987}. For details of other state-dependent versions of the Bell-Kochen-Specker theorem see \citet{brown:1992}, section 2(iv), where it is argued that a result due to \citet{belinfante:1973} can be construed as such. \label{HomeSengupta}}

But for our present purposes, the relevant question that Bell raised in his review paper was whether the action-at-a-distance built into the \citeyear{bohm:1952} Bohm pilot wave theory was characteristic of all hidden variable theories. As Bell noted, in the Bohm theory, in the case of measurements on a pair of separated entangled systems, `the disposition of one piece of apparatus affects the results obtained with a distant piece'.
Part of the answer to the question was provided in Bell's 1964 paper, amusingly published two years earlier. (Advocates of the retrocausal `free will' loophole take heart.) Bell showed that any deterministic theory in which the hidden variable is associated with the quantum system alone (i.e. supplements the system's quantum state vector) must display action-at-a-distance of a Bohmian nature, if it is to be consistent with the predictions of standard quantum mechanics. Specifically, what must be violated, in the case of spatially separated spin-1/2 systems in the singlet state, is that \textit{The result of a Stern-Gerlach measurement of a spin component on either system does not depend on the setting of the magnet for the other system.}
(The more general claim that in a \textit{deterministic} hidden variable theory, the predicted outcome of a measurement of an observable on one of a pair of entangled systems does not depend on how a distant piece of equipment designed to measure any observable on the other system is set up, will be referred to as the \textit{1964 locality assumption}.)

Bell's 1964 result was ground-breaking in itself. But it is fair to say that it was only part of the way to an answer to the specific question posed in the review paper, because the Bohm theory generally involves in its deterministic algorithm hidden variables associated with both the object system and the apparatus. Bell was to look at generalisations of this kind of theory in 1971, and unwittingly anticipated the later notion of local causality, as we shall see shortly.\footnote{It is noteworthy that recently a third aspect of de Broglie--Bohm theory has been shown to be universal for hidden variable theories: that the quantum state can not be entirely `epistemic' in nature. See \citet{PBR} and \citet{BCLM}.}

Bell started his 1964 paper by recalling that the 1935 Einstein-Podolsky-Rosen (EPR) argument demonstrates, using the strict, or perfect correlations involved in entangled systems, that if action-at-a-distance is to be avoided, the standard quantum mechanical state could not be a complete description of the systems in question.\footnote{We disagree with a recent claim that the EPR argument presupposed more than standard quantum mechanics (with measurement induced collapse) and locality, the `more' being counterfactual definiteness; see \citet{ZukowskiandBrukner:2014}.}  In particular Bell seems to have accepted that the perfect anticorrelations for parallel spin components in the singlet state of spin-1/2 systems imply that a deterministic underpinning (`causality') is necessary if locality in this sense is to prevail, and that this is the reason why a local \textit{deterministic} hidden variable theory is the focus of his 1964 paper.\footnote{See \citet{maudlin:what} for making the case that Bell viewed determinism as a strict consequence of the EPR argument based on locality (no action-at-a-distance). Certainly after 1964, Bell was explicit: `It is important to note that to the limited degree to which \textit{determinism} played a role in the EPR argument, it is not assumed but \textit{inferred}.' \citet[p.143]{bell:bertlmann}, original emphasis. However, disagreement has arisen in the literature as to the precise logic of Bell's 1964 theorem, with particular reference to the role of determinism. Some commentators see determinism as one of the assumptions of the theorem (view 1), while others see it as a consequence of the assumptions, which include the existence of perfect (anti)correlations. Clearly, if the former position is correct, then the empirical violation of the Bell inequality implies \textit{either} indeterminism \textit{or} nonlocality (in the 1964 sense of the term). A recent careful textual analysis of Bell's writings in the context of this debate is due to \citet{Wiseman:2014}, who provides grounds for thinking that in later life Bell's own reading of his 1964 logic---in line with view 2---is questionable. Note that nothing in our paper hangs on this debate.} Bell showed that a local deterministic hidden variable theory can easily be constructed to account for such correlations, so he was fatefully drawn to investigate the weaker correlations between non-parallel spin components on the distant systems: literally the EPR--Bohm scenario with a twist. 

In so far as Bell was to drop the deterministic requirement in his later papers, the connection with the EPR argument became less tangible. And this had to do with the fact that in its most general form, the derivation of Bell-type inequalities based on some version of the locality condition transcends quantum mechanics. It does not have to appeal to the existence of perfect EPR--Bohm (anti-)correlations between distant systems. We shall return to this point below, but in the meantime it is worth recalling the conclusion Bell draws from his 1964 theorem.
\begin{quotation}
In a theory in which parameters are added to quantum mechanics to determine the results of individual measurements, without changing the statistical predictions, there must be a mechanism whereby the setting of one measurement device can influence the reading of another instrument, however remote. Moreover, the signal involved must propagate instantaneously, so that such a theory could not be Lorentz invariant. \citep{bell:1964}
\end{quotation}
The conclusions in Bell's 1976 and 1990 papers are much more nuanced. There, as we shall see, there is no clear-cut claim of conflict with special relativity, largely because of explicit recognition of what is now widely known as the no-signalling theorem of quantum mechanics, 
which of course holds at the statistical level and hence is insensitive to the existence of any nonlocal hidden substratum.\footnote{The no-signalling theorem in quantum mechanics only became prominent in the literature in the late 1970s, though its first enunciation effectively goes back to David Bohm's \citeyear{bohm:book} book \textit{Quantum Theory}. (Some historical details regarding the no-signalling theorem are found in \citet{timpsonandbrown:2002}, footnote 12; Bell's 1976 discussion is regrettably overlooked.)} It is interesting though that a degree of caution, if not skepticism, about the theorem emerges from Bell's remarks (see below).

\section{The legacy of Einstein-Podolsky-Rosen}

We have seen that for Bell in 1964, the EPR argument was crucial in setting up the conditions that lead to an inequality which is violated by quantum mechanics. Einstein's own conviction that locality (absence of action-at-a-distance) and the completeness of quantum mechanics are incompatible actually predate the 1935 collaboration with Podolosky and Rosen. In the 1927 Solvay conference, Einstein used the single slit scenario to argue that detection of the particle at one point on the hemispherical measuring instrument means that all other points must instantaneously know not to detect, despite the wavefunction of the particle having finite value at all such points prior to detection.\footnote{\citet{brown:1981} and \citet{norsen:2005}.} In fact, given the completeness assumption, measurement-induced collapse of the wavefunction involves action-at-a-distance: \textit{one does not need entanglement and EPR correlations to drive the nonlocality lesson home given such non-unitary processes.} 

Unless one adopts something like Niels Bohr's philosophy. Here, the wavefunction/state vector is thought to be the complete description of the quantum system but somehow not in itself a physical object, and collapse (if one goes so far as to describe in quantum mechanics what goes on the measurement process, which Bohr did not) is thought not to represent a physical process.\footnote{See, for example, \citet[p.53]{bell:speakable}.} So the EPR argument is best seen, like Newton's bucket thought experiment, as polemical in nature. Newton used the bucket to strike at the heart of Descartes' theory that for any body real motion is defined relative to the bodies immediately contiguous to it, themselves being taken to be at rest.\footnote{See \citet[pp.623--628]{barbour:discovery}.} By cleverly exploiting \textit{distant} correlations, Einstein hoped to bypass the feature of quantum mechanics that Bohr commonly used in defence of the claim that quantum mechanics is complete notwithstanding its statistical character, namely, the ineradicable \textit{local} disturbance on the system caused by measurement. This aspect of the EPR argument was appreciated by Bell, who in 1981 devastatingly  exposed the obscurity of Bohr's 1935 response to the argument.\footnote{See \citet{bell:bertlmann}, Appendix 1.}

Be that as it may, it is well known that Einstein was unhappy with the way Podolsky had organised the argument, and in letters to Schr\"{o}dinger had vented his frustration that the basic lesson had been `smothered by the formalism'.  Details concerning the form of the argument that Einstein preferred will not be rehearsed here\footnote{A useful account of Einstein's criticisms of the EPR paper is found in \citet{fine:shaky}, chapters 3 and 5; see also \citet{maudlin:what} and \citet{timpsonandbrown:2002} for further discussion of Einstein's overarching view on locality.}, yet it  is worth mentioning the thought experiment known as `Einstein's boxes'.\footnote{This terminology is due to \citet[p.37]{fine:shaky}; a fuller discussion of the history of the thought experiment is found in \citet{norsen:2005}.} Writing to Schr\"{o}dinger in 1935, Einstein imagined a box with a single classical particle inside. The box is then divided into two and each half-box spatially separated. The particle by chance ends up in one of them, and Einstein used this scenario to explain what he meant by an `incomplete' description of the particle and what he meant by the `separation principle' (locality), all with a view to better articulating the quantum EPR scenario involving entanglement. But it was de Broglie in 1964\footnote{See \citet{deBroglie:1964} and \citet{norsen:2005}.} who considered the analogue of the boxes scenario with a quantum particle, its wavefunction now in a superposition of components in each half-box prior to their opening. Given the strict anticorrelation between the outcomes of opening the boxes, this gedanken experiment has all the features required for an EPR-type argument inferring nonlocality (action-at-a-distance) from completeness of the wave functional description, but without entanglement. It is a variant of Einstein's 1927 single slit diffraction argument, and both arguments will feature below.

\section{Local causality}\label{localcausalitysection}

Suppose now we step back from quantum mechanics and consider some hypothetical probabilistic theory involving microscopic systems. Suppose the probabilities referring to measurement outcomes in the theory are understood to be irreducible, the dynamics of the relevant processes being intrinsically stochastic.\footnote{Technically what this means is that  the theory allows no dispersion-free pure ensembles and yet is assumed to be `statistically complete'. Recall that pure states are those which cannot be expressed as a convex combination of other distinct states. Statistical completeness entails that arbitrary ensembles of the joint system-apparatus which are pure---i.e. all joint systems therein share the same pure state---are homogeneous in the sense of von Neumann. A detailed analysis of this notion of statistical completeness, which is distinct from the EPR notion of completeness, is found in \citet{elbyetal:1993}.} Imagine further that a pure ensemble of spatially separated bipartite systems along with measurement devices can be prepared in which by hypothesis no interactions are taking place between them. Would we not expect the statistical outcomes of the measurement events on the separated systems to be independent, i.e. for there to be no correlations?

Something akin to this no-correlations condition was introduced into Bell's writings explicitly in 1976. We shall examine its formal representation, as well as its (dubious) connection with Hans Reichenbach's famous Common Cause Principle \citep{reichenbach:commoncause}, in Section~\ref{unpacking} below. In the meantime a little history may be useful.

In his second paper on nonlocality, published in \citeyear{bell:introtohv}, Bell considered a class of hidden variable theories explicitly modelled on the 1952 Bohm theory, in which the predictions of measurement outcomes depend deterministically on hidden (hence uncontrollable) variables associated both with the object system and the apparatus. Averaging over the latter, Bell noted that the theory now took the form of an \textit{indeterministic} hidden variable theory for measurements occurring on the object system. In his derivation of the \citeyear{CHSH} Clauser--Horne--Shimony--Holt (CHSH) inequality for this stochastic theory involving pairs of distant systems, Bell took as his locality condition that these average predictions on one system did not depend on the controllable setting of the distant measurement device. But another assumption was hidden in the derivation: that such averages for pairs of simultaneous measurements on the distant systems factorize.\footnote{See \citet[p.1385]{brownandsvetlichny:1990}, and \citet[\S {II}]{brown:1991}).}   It is true that this is a consequence of the the locality of the background deterministic theory,\footnote{Cf. \citet[\S {II}]{brown:1991}.} but it was a foretaste of what was to come in Bell's thinking about his theorem.\footnote{Bell noted that the 1971 locality condition `Clearly...is appropriate also for \textit{indeterminism} with a certain local character.' \citep[fn.10]{bell:introtohv} original emphasis.}

Note that Bell himself was aware that if one added to his 1971 derivation the existence of strict EPR--Bohm anticorrelations in the spherically symmetric spin singlet state, then his original 1964 version of the inequality is obtained from that of CHSH. More to the point, Bell was conscious that allowing for the strict anticorrelations meant that the apparatus hidden variables can play no role in the predictions, so that even the apparent indeterminism of the averaged theory is spurious. But one sees in this 1971 paper that Bell was starting to follow CHSH in considering generalised local theories as the proper background to the Bell theorem, in which no constraints on correlations exist other than those imposed by locality. 

The first clear articulation of this program for the case of genuinely stochastic (indeterministic) theories was due to John Clauser and Michael Horne in \citeyear{ch:1974}. In defining locality in this context these authors explicitly distinguished between two predictive constraints: independence of the distant apparatus setting (sometimes called \textit{parameter independence}) and the absence of correlations conditional on specification of the pure state of the pair of systems (\textit{conditional outcome independence}, or \textit{Jarrett completeness}\footnote{The widespread appreciation that two distinct assumptions are at play in the early versions of the stochastic version of the Bell theorem is due in good part to the work of \citet{jarrett:1984}.}).  The subtle business of how to motivate these constraints, and the connection, if any, with the principle of no action-at-a-distance, will be discussed below in Section~\ref{unpacking}. In \citeyear{suppeszanotti}, Suppes and Zanotti proved in a general way what Bell noted in his 1971 paper, namely that if one imposes these two `locality' constraints for a stochastic theory of pairs of systems, and moreover postulates the existence of perfect (anti)correlations of the kind associated with the spin singlet state, then the theory is reduced to a deterministic one \citep{suppeszanotti}.\footnote{The same situation holds for the triplet state for spin-$\frac{1}{2}$ systems despite its lack of spherical symmetry. This is because for any spin component on one subsystem, there is a spin component on the other with which it is perfectly correlated (though this is not in general the component anti-parallel to the first as in the case of the singlet state).}

Suppose only conditional outcome independence is assumed, and not parameter independence. Then the work of Suppes and Zanotti (1976) implies in the case of the spin singlet state that the marginal probabilities for values of the spin components are zero or one in every case where the spin-devices are parallel-oriented. So there will be elements of reality at one wing of the experiment which can be brought in and out of existence by varying the orientation of the device at the other wing.
In this case, the joint locality condition reverts to the prohibition of action-at-a-distance. However, it is important to bear in mind at this point that an implicit, but highly non-trivial assumption is being made in the Suppes and Zanotti argument, as well as in Bell's own application of perfect (anti)correlations in order to restore determinism: that in each measurement process, one and only one outcome is realised. We return to this issue in Section~\ref{Everett} below.

In 1976 there also appears for the first time in Bell's writings both the explicit notion and terminology of `local causality' associated with stochastic theories of physical reality.\footnote{Confusingly, at times Bell would use the term `local causality' when he explicitly meant no action-at-a-distance in the context of the EPR argument, as in \citet[p.143]{bell:bertlmann}.} As he was famously to articulate in his later \citeyear{bell:against} `Against measurement', Bell was deeply suspicious of the cavalier way such common but non-fundamental notions as `observable', `system' and `measurement'  are often used in quantum mechanics. In his 1976  paper on `local beables', Bell defines the factorizability condition not in terms of measurement outcomes on bipartite systems, but in terms of the beables in spacelike separated regions of space-time, conditional on the complete specification of all the beables belonging to the overlap of the backward light cones of these regions. There is no splitting of the local causality condition into anything like parameter independence and conditional outcome independence, because measurement processes do not explicitly figure in the analysis. Two inequalities are derived in Bell's 1976 paper. The first involves nothing but beables and their correlations; the second---a variation of the CHSH inequality---emerges from the first when `in comparison with quantum mechanics', some beables are interpreted as controllable variables specifying the experimental set-up, and some are `either hidden or irrelevant' and averaged over as in the 1971 paper.

\section{The 1976 paper}\label{the1976paper}

There are several remarkable features of Bell's 1976 paper. 

\begin{enumerate}

\item More than ever before in Bell's writings, \textit{his theorem stands alone from quantum mechanics.} In contradistinction to his 1971 paper, Bell starts by considering a loosely defined but \textit{genuinely} indeterministic theory of nature, and the derivation of the (first) inequality conspicuously holds without any reference to quantum features, and depends only on the requirement of local causality. 
\begin{quote}We would like to form some [notion] of local causality in theories which are not deterministic, in which the correlations prescribed by the theory, for the beables, are weaker. \citep[p.53]{bell:tlb}
\end{quote}

\item For the first time, Bell states that `ordinary quantum mechanics, even the relativistic quantum field theory, is not locally causal'. \textit{The claim does not depend on his theorem. It does not depend on the violation of a Bell-type inequality: it does not even depend on entanglement in the usual sense.} The argument Bell gives concerns a single radioactive decay process and several spatially separated detectors; it is analogous to Einstein's 1927 argument involving single slit diffraction, and to the Einstein-de Broglie boxes thought experiment, both referred to earlier. (This aspect of Bell's reasoning has been little appreciated in the literature, and it was partially lost sight of in his final 1990 paper.)

\item Bell importantly qualifies the significance of local causality in the light of the no-signalling theorem, which makes it first appearance in his work, and which itself is qualified in terms of its `human' origins. This hesitancy is repeated in 1990, as we see in the following section.

\item When Bell does apply his theorem to quantum mechanics, it is in the context of a putative locally causal, indeterministic hidden variable `completion' of the theory. But now there is no mention of the EPR-Bohm (anti)correlations: the legacy of EPR, so vital in the 1964 paper, has been put aside. 
\end{enumerate}

Let us ruminate a little on this last point. To ignore perfect EPR--Bohm (anti)correlations is not to deny their existence. But if they exist, then, as we have seen, in the usual setting, stochasticity collapses into determinism and the meaning of local causality reverts simply to the prohibition of action-at-a-distance. It is understandable then that Maudlin in 2010 construes the legacy of Bell's work to have been the demonstration that standard quantum mechanics is nonlocal in the original 1964 sense, and that the proof does \textit{not} presuppose the existence of hidden variables, or the related condition of `counterfactual definiteness', as long as EPR--Bohm correlations are assumed and the possibility of  Everettian branching is ruled out (see below). Rather, the existence of a deterministic substratum (and hence counterfactual definiteness) is \textit{inferred} in the proof, and locality further constrains the  phenomenological correlations so as to satisfy a Bell inequality, which quantum mechanics violates in certain scenarios involving entangled states.\footnote{\citet{maudlin:what}. Maudlin is one of those defending view 2 mentioned in footnote~\ref{HomeSengupta}  above, namely that determinism is not an assumption in Bell's original 1964 theorem. Again, we emphasise that we are interested here in a certain reading of the theorem, not whether Bell himself adopted it in 1964.}

This view of things does not seem to accord with Bell's thinking from 1976 on. As we have seen, the demonstration that quantum mechanics violates the new condition of local causality is fairly elementary and needs no appeal to the Bell theorem or even entanglement. The Bell theorem itself refers to locally causal stochastic theories generally, and when it is applied to the specific case of quantum mechanics, it is to some \textit{stochastic} `completion' of the theory which can have no justification in terms of the EPR argument. In 1981, in his famous paper `Bertlemann's socks and the nature of reality', Bell made clear what his position on the EPR--Bohm anticorrelations came to be:
\begin{quote}
Some residual imperfection of the set-up would spoil the perfect anticorrelations ... So in the more sophisticated argument we will avoid any hypothesis of perfection.

It was only in the context of perfect correlation (or anticorrelation) that \textit{determinism} could be inferred for the relation of observation results to preexisting particle properties (for any indeterminism would have spoiled the correlation). Despite my insistence that determinism was inferred rather than assumed, you might still suspect somehow that it is a preoccupation with determinism that creates the problem. Note well then that the following argument [derivation of the CHSH inequality based on local causality] makes no mention whatever of determinism. \citep{bell:bertlmann} emphasis in original.
\end{quote}
Granted, \textit{perfect} (anti)correlations are operationally inaccessible; indeed this was the motivation for the CHSH inequality. This inequality is testable in a way Bell's original 1964 inequality is not. But we should not lose sight of the fact that every experimental corroboration of the general correlation predictions in quantum mechanics for entangled states provides indirect evidence for the perfect (anti)correlations in the EPR-Bohm scenario, which in turn are connected with the fact that the spin singlet state has zero total spin angular momentum. It is noteworthy that the perfect (anti)correlations reappear in Bell's 1990 paper, as seen below. 

We are thus led to the view that from 1976 on in Bell's writings, when he refers to a putative locally causal, stochastic, hidden variable `completion' of quantum mechanics, this theory should be understood in the following sense.  \textit{Either it is a truncated theory restricted to certain non-parallel settings, or one incorporating an approximation to quantum mechanics whose correlation predictions concur with those of quantum mechanics for parallel settings to within experimental error (and so the theory may depend on the chosen experimental set-up).}  Otherwise, the theory is not stochastic at all, as we saw in Section~\ref{localcausalitysection}.

\section{Local causality, no-signalling and relativity}

The factorizability condition related to local causality first introduced by Bell in 1976 was itself to undergo a minor change in his later writings. He came to realise that it is unnecessary to conditionalize on complete specification of beables in an infinite space-time region, and a `simpler' version of the condition involving beables in finite regions appeared informally in a footnote in \citet{bell:EPREPW}.\footnote{\citet[fn. 7]{bell:EPREPW}. In his Preface to the first edition of \textit{Speakable}, Bell expressed regret that this slimmer version, which he had used in talks, had not been introduced earlier in his papers; see \citet[p.xii]{bell:speakable}. There is also a reminder in the Preface that `If local causality in some theory is to be examined, then one must decide which of the many mathematical entities that appear are supposed to be real, and really here rather than there.'\textit{ibid}.} A more systematic rendering of this `simpler' version appeared in Bell's final 1990 paper `La nouvelle cuisine', which first appeared in the second edition of \textit{Speakable and Unspeakable in Quantum Mechanics}. 

The details need not detain us yet. But note that in the 1990 paper, when Bell discusses the reason for the violation of local causality by `ordinary quantum mechanics', as in the 1976 paper he makes no to appeal to the violation of a Bell-type inequality and hence to the Bell theorem, but now he does refer to entanglement and in particular an optical version of the EPR--Bohm scenario with perfect correlations. (Compare with point 2 in the previous section.)

Does the fact that quantum mechanics violates local causality mean it is inconsistent with special relativity? This is a question over which Bell wavered for a number of years, and in attempting to understand his position it is important to distinguish between `ordinary' quantum mechanics and the hypothetical `deeper' level of hidden variables. 

Recall that in 1964, Bell categorically concluded that any deterministic hidden variable theory consistent with the standard quantum mechanical predictions could not be Lorentz covariant, though in 1976 he was more nuanced about the violation of local causality. In 1984 he wrote:
\begin{quotation}
For me then this is the real problem with quantum theory: the apparently essential conflict between any sharp formulation and fundamental relativity. That is to say, we have an apparent incompatibility, at the deepest level, between the two fundamental pillars of contemporary theory ...\citep{bell:speakablepaper}
\end{quotation}
Note the qualifying adjectives `apparent' and `fundamental'. In his influential essay of 1976, `How to teach special relativity' \citep{bell:howtoteach}, Bell advocated a dynamical Lorentzian pedagogy in relation to the explanation of kinematic effects such as length contraction and time dilation. He stressed that one is not thereby committed to a Lorentzian `philosophy' involving a privileged inertial frame, as opposed to Einstein's austere philosophy which rejects the existence of such a frame. However, Bell also emphasised that `The facts of physics do not oblige us to accept one philosophy rather than the other.' When he was asked in a recorded interview, first published in 1986, how he might respond to the possible existence of `faster-than-light signalling' in the Aspect experiment, Bell returned to this theme.
\begin{quote}
... I would say that the cheapest resolution is something like going back to relativity as it was before Einstein, when people like Lorentz and Poincar\'{e} thought that there was an aether---a preferred frame of reference---but that our measuring instruments were distorted by motion in such a way that we could not detect motion through the aether. Now, in that way you can imagine that there is a preferred frame of reference, and in this preferred frame of reference things do go faster than light. ... Behind the apparent Lorentz invariance of the phenomena, there is a deeper level which is not Lorentz invariant ... [This] pre-Einstein position of Lorentz and Poincar\'{e}, Larmor and Fitzgerald (\textit{sic}), was perfectly coherent, and \textit{is not inconsistent with relativity theory}. \citep{bell:ghost} our emphasis.
\end{quote}
This is a very different tone to the one at the end of Bell's 1964 paper, in which the failure of Lorentz covariance in any deterministic hidden variable theory is announced without any redemptive features. In 1986 he is adopting the perfectly defensible view that special relativity strictly only holds for physics that has phenomenological consequences, i.e. that is not `hidden'. Note that Bell does not clarify here \textit{what} goes faster than light  in the `deeper level' in quantum mechanics. Given the context of the discussion, it seems that something like the action-at-a-distance in the Bohm theory is what Bell had in mind. But the question is more pressing if the issue is violation of local causality, when it is just `ordinary' quantum mechanics or quantum field theory that is in question.

In his `La nouvelle cuisine' paper, Bell harks back to his 1976 discussion of the nonlocality theorem and provides a particularly careful demarcation between the condition of local causality and that of `no-superluminal-signalling'. In connection with the latter, Bell investigates the case of external interventions in local relativistic quantum field theory, and concludes, unsurprisingly, that the statistical predictions are insensitive to the introduction of external fields outside the backward light cone of the relevant `observables'. But he is worried by what is truly meant by the notion of `external' intervention. 
At the end of the paper he returns to the no-signalling theorem in standard quantum mechanics, and he similarly expresses concerns about its fundamentality, just as he had in 1976. He is worried, as always, about the vagueness or lack of conceptual sharpness in the theorem, which results from the anthropocentric element lurking behind the notions therein of measurement and preparation, and indeed this issue appears to be the lingering concern in the conclusion to the 1990 paper. 
\textit{There is no categorical statement in the paper that violation of local causality is inconsistent with special relativity.}

\section{Motivating local causality}

In `La nouvelle cuisine', Bell emphasised that even if a theory is well-behaved relativistically in the strict sense of allowing for no superluminal signalling, this is not enough to satisfy his causal intuitions. What then is the motivation for local causality? We need, at last, to look in detail at the definition and significance of the factorizability condition as given in this 1990 paper. Here is what Bell says:
\begin{figure}
\begin{center}
      \includegraphics[scale=0.60]{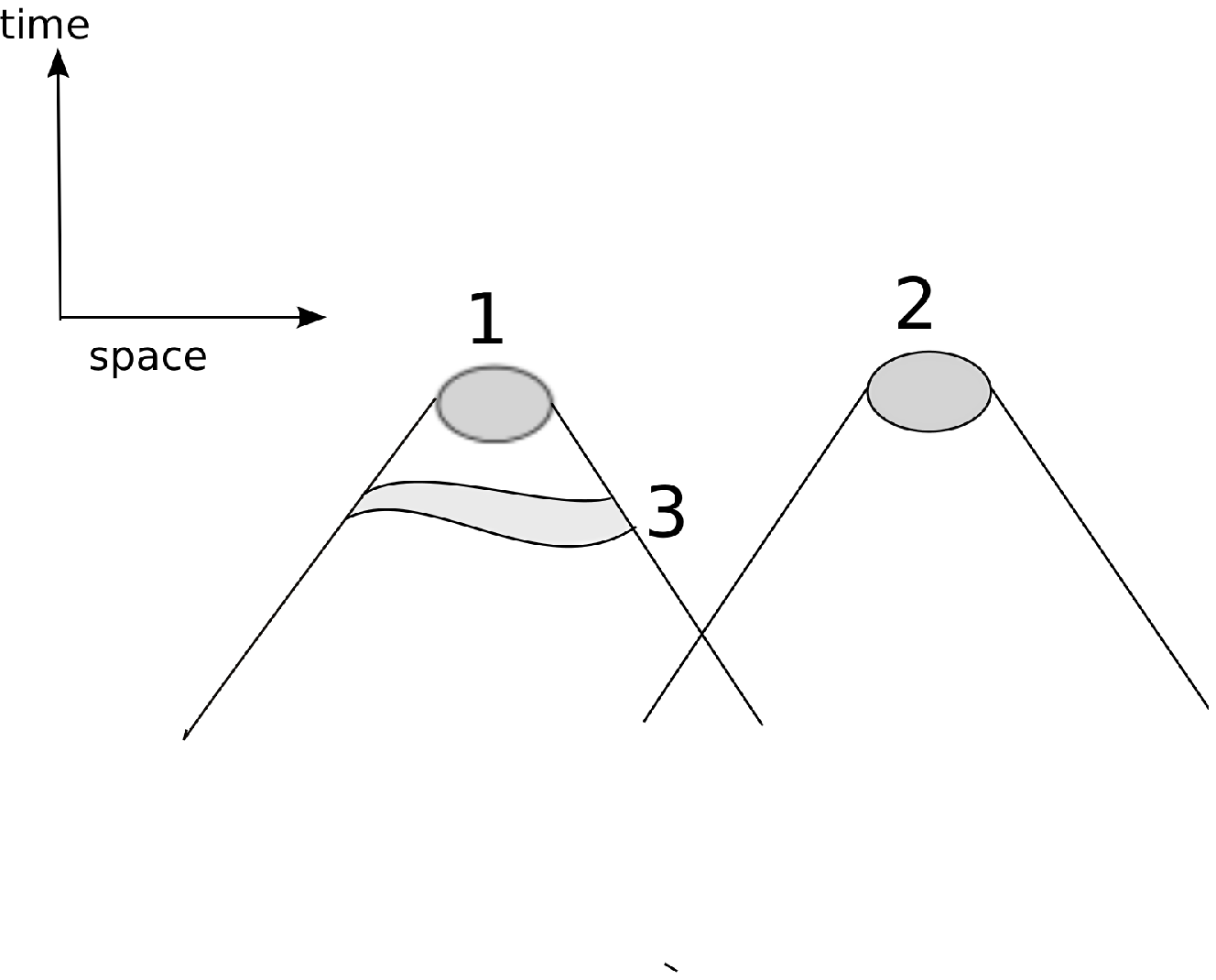}
  \end{center}
\caption{Spacetime regions involved in Bell's 1990 statement of local causality}
\end{figure}
\begin{quote}
A theory will be said to be locally causal if the probabilities attached to values of local beables in a space-time region 1 are unaltered by specification of values of local beables in a space-like separated region 2, when what happens in the backward light cone of 1 is already sufficiently specified, for example by a full specification of local beables in a space-time region 3... \citep[pp.239--40]{bell:nouvelle}
\end{quote}

What is particularly relevant for our purposes is the intuition behind local causality that Bell provided in 1990:
\begin{quotation}
The direct causes (and effects) are near by, and even the indirect causes (and effects) are no further away than permitted by the velocity of light.\citep[p.239]{bell:nouvelle}
\end{quotation}
It is important for our subsequent discussion to highlight some key assumptions involved here. The first is the very existence of causal processes, which are normally taken to be time-asymmetric: causal influences propagate into the future, not the past. (Note the reference to backward light cones in the first quote.) An awkwardness arises because there is nothing in the postulates of special relativity that picks out an arrow of time. (Whether the situation is any different in quantum mechanics is clearly an interpretation-dependent issue.) At any rate, operationally the notion of cause had to be cashed out by Bell in probabilistic terms; he conceded that the factorizability condition associated with local causality makes no explicit reference to causes.  `Note ... that our definition of locally causal theories, although motivated by talk of ``cause'' and ``effect'', does not in the end explicitly involve these rather vague notions.' \citep[p.240]{bell:nouvelle} 

Note that the introduction of probabilities introduces further awkwardness into the picture. Special relativity is based on the claim that the fundamental equations governing the non-gravitational interactions are Lorentz covariant. Such equations are normally assumed to be deterministic. How irreducible probabilities are supposed to inhabit a relativistic world is a difficult technical matter, which Bell, perhaps wisely, overlooks. And it is worth recognising at this point that unless probabilities in this context are understood themselves to be objective elements of reality, things go awry. There is an old view, by no means consensual of course, that probability in physics in general does not represent an objective element of reality.\footnote{A version of this view, in which probability is  `a  numerical expression of human ignorance', has recently been defended by \citet{tipler:2014} for example. See also \citet{cfs:certainty,fuchs:greeks} and Healey \citeyear{healey:eps2012}, and this volume, for various developments of related ideas.} If this were true, the representation of purportedly objective causal facts by constraints on probabilities becomes problematic and the status of factorizability somewhat questionable. In particular, the connection with the clear-cut condition of no action-at-a-distance involved in Bell's 1964 paper becomes tenuous. This concern might seem misplaced in the context of an indeterministic theory, but what about a deterministic theory with intrinsic unpredictability, as in the Everett picture (see next section)?

The second (widely-held) assumption Bell is making is that the null cone structure of Minkowski space-time represents the boundaries of causal connectibility. The kinematics of special relativity imply that the speed of light is invariant across inertial frames, and the usual dynamical postulates (over and above Lorentz covariance) concerning the connection between energy, momentum and velocity imply that massive bodies cannot be accelerated up to the speed of light, let alone beyond. But whether tachyonic signals are inconsistent with special relativity is a more delicate matter, and Bell was fully aware of this. As he put it: `What we have to do is add to the laws of relativity some responsible causal structure. ... we require [causal chains] ...  to go slower than light in any [inertial] frame of reference.' \citep{bell:nouvelle} Bell was effectively---and plausibly---assuming that tachyonic signals do not exist. But the justification Bell gave in section 4 of the Nouvelle Cuisine paper is amusing. It involves considering the `perfect tachyon crime' involving a gunman firing a tachyon gun. The movement of the murderer can be set up in such a way that according to the description of the deed relative to the rest frame of the victim, and the courts of justice, the trigger is pulled after the death of the victim! The ensuing `relativity of morality' should be ruled out by the laws of nature, suggests Bell. This injunction is obviously high-handed---why should Nature care about our moral qualms? Playfully or otherwise, Bell is insisting on first principles that superluminal signalling be banished from Minkoswksi space-time, and as we mentioned in the last section, he goes on to show in the 1990 paper that quantum theory does not violate this injunction. But, to repeat, there is still a distance between no-signalling and local causality, particularly in the case where the candidate indeterministic theory is acting as a completion of ordinary quantum mechanics, as will be seen in the next section. 

Bell emphasised that the factorizability condition is not to be seen `as the \textit{formulation} of local causality, but as a consequence thereof.' He was very concerned that in providing  a clean mathematical consequence of local causality, one is `likely to throw out the baby with the bathwater'. 
Indeed, careful analysis of the task that presents itself in the attempt to translate Bell's causal intuitions into well-defined probabilistic form shows how subtle the matter is. This has become especially clear through the recent work of \citet{Norsen:2011}, and \citet{seevinckuffink:bathwater}. However, we will stick to the specific form of factorizability Bell gave in the Nouvelle Cuisine paper as applicable to EPR--Bell experiments, where \textit{the candidate indeterministic theory is now understood to be a completion of ordinary quantum mechanics} (subject to the qualification mentioned at the end of Section~\ref{the1976paper}). Here, Bell extended the space-time region 3 in the figure above to cross the backward light cones of both regions 1 and 2 where they no longer overlap. In this extended region 3, Bell denoted by the symbol $c$ the values of any number of variables describing the experimental set up as given in ordinary quantum mechanics, except the final settings (of magnets, polarisers, or the like) that are determined in regions 1 and 2. Presumably included in $c$ is the entangled quantum state. Also defined in the extended region 3, and denoted by $\lambda$, is `any number of additional complementary variables needed to complete quantum mechanics in the way envisaged by EPR.' Note that they too are not assumed to be spatially separable. Together $c$ and $\lambda$ give a `complete specification of at least those parts of 3 blocking the two backward light cones'. Then Bell's specific 1990 factorizability condition is 
 \begin{equation}
P(A, B| a, b, c, \lambda) = P(A|a, c, \lambda)P(B| b, c, \lambda)
\end{equation}
where $P(A, B| a, b, c, \lambda)$ is the joint probability in the candidate theory for outcome $A$ associated with a binary random variable in an experiment in region $1$ with setting (free variable) $a$, and outcome $B$ in an experiment in region $2$ with setting $b$, conditional on $c$ and $\lambda$. The factors in the product on the RHS are the respective conditional marginal probabilities.

\section{Unpacking factorizability}\label{unpacking}

As we have mentioned, ever since the 1974 work of Clauser and Horne, reinforced by that of Jarrett and also of Shimony (e.g., \citet{shimony:controllable}), it has been regarded as conceptually helpful to separate any such factorizability condition into two components:
\begin{subequations}
\begin{align} 
P(A|a, b, c, \lambda) = P(A|a, c, \lambda), \\
P(B|a, b, c, \lambda) = P(B|b, c, \lambda),
\end{align}
\end{subequations}
and
\begin{subequations}
\begin{align} 
P(A|B, a, b, c, \lambda) = P(A|a, b, c, \lambda), \\
P(B|A, a, b, c, \lambda) = P(B|a, b, c, \lambda).
\end{align}
\end{subequations}
Equations (2) indicate that the marginal probabilities do not depend on the settings of the distant piece of apparatus. This `parameter independence' assumption (`functional sufficiency' in the language of Seevinck and Uffink) is the probabilistic analogue of the locality assumption in Bell's 1964 paper. Equations (3) known, as we have said, as `conditional outcome independence' or `Jarrett completeness' (`statistical sufficiency' for Seevinck and Uffink) rule out correlations conditional on the combination of $c$ and $\lambda$. A great deal of discussion has of course taken place since the 1970s about these conditions, and indeed their non-uniqueness, but let us note the claim by Seevinck and Uffink that each condition has the same motivation as the factorizability condition (1) itself: `Both [conditions] are a consequence of local causality, and the appeal to notions of locality and causality used in implementing the functional and statistical sufficiency are just the same ...'' \citep[p.12]{seevinckuffink:bathwater}

But now that the candidate indeterministic theory is one into which quantum mechanics is embeddable,  the specification of $\lambda$ involves hidden variables. Wouldn't doubt be thrown in this case on the plausibility of parameter independence (2), in particular, given Bell's own doubts whether what is `hidden' can ever threaten relativity?  Maybe, but as we have seen in the previous section, the issue is not just consistency with relativity, whether in the sense of requiring Lorentz covariance or in the sense of precluding tachyonic signalling. As regards outcome independence (3), ordinary quantum mechanics (in which $\lambda$ is the empty set) violates it, but again no threat to a `responsible causal structure' containing a ban on tachyonic signals automatically arises as a result. 

Let us consider this condition (3) in more detail. It is often regarded as a special case (within a relativistic space-time) of Reichenbach's \citeyear{reichenbach:commoncause} Common Cause Principle,\footnote{See for example \citet{vanFraassen:1982}, \citet{Butterfield:1989,Butterfield:1992,Butterfield:2007}, \citet{Henson:onticindefiniteness}.} which is based on the general notion that if events $X$ and $Y$ are correlated, then either $X(Y)$ is the cause of $Y(X)$ or there is a common cause in the past \citep[\S 19]{reichenbach:commoncause}. Conditional on the common cause, the correlations between $X$ and $Y$ must disappear (the so-called `screening-off' condition). But although this principle is a useful heuristic in daily life, Reichenbach's own writings on the matter should make us pause before applying it to microphenomena (whether classical or quantum) involving a small number of systems. What is amply clear in Reichenbach's 1956 discussion of the principle is that it holds for `macrostatistics', that is to say systems sufficiently complex to display entropic behaviour. For Reichenbach, the notions of cause and effect only make sense in macro-phenomena displaying quasi-irreversible behaviour: `The distinction between cause and effect is revealed to be a matter of entropy and to coincide with the difference between past and future.' \citep[p. 155]{reichenbach:commoncause} There is of course irreversibility associated with the measurement processes themselves in an EPR--Bell experiment, but it may not be enough to resolve the conundrum in Reichenbach's own terms, although he was not entirely consistent on the whole matter.\footnote{At the end of his 1956 book, Reichenbach asserted that quantum mechanics---\textit{even at the statistical level}---no more introduces a fundamental arrow of time than does classical mechanics (p. 211). As for probabilities themselves, Reichenbach defined them in terms of relative frequencies, so they are time symmetric in his book. It is then hard to see how he could apply his common cause principle to quantum phenomena involving small numbers of systems, and yet he seems to have done so! Reichenbach states in the book that the statistics of pairs of identical bosonic particles engender `causal anomalies'---i.e. violate his principle---unless the assignment of physical identity to the particles is given up (p. 234).} 

To the extent that Reichenbach's common cause principle \textit{is} cogently applicable to Bell-type experiments, there is arguably a missing element in Reichenbach's analysis, which was largely motivated by common-place correlations involving familiar day-to-day phenomena.
In the case of a strictly indeterministic microphysics of the kind Bell considered from 1976 on, the motivation for conditional outcome independence is arguably stronger than anything Reichenbach had in mind. Bell was envisaging  outcomes of measurements in regions 1 and 2 which are \textit{intrinsically} random, associated with  \textit{irreducible} probabilities. \textit{What kind of explanation can there be for correlations between separate, strictly random processes?} Note that the puzzle does not require reference to space-like separations defined relative to the light cone structure; it would presumably arise even in Galilean space-time for simultaneous events as long as the dynamics in the candidate indeterministic theory did not allow for instantaneous interactions between separated systems.  The profound oddity of such correlations---let us call it the \textit{randomness problem}---goes hand in hand with the natural view that probabilities in strictly indeterministic theories are, \textit{pace} Tipler, objective. (We shall return to the randomness problem in Section~\ref{everettiananalysis}.) 

Bell himself did not put things quite this way. A noteworthy element of his surprisingly tentative, but always honest, thinking in the Nouvelle Cuisine paper comes in the Conclusion: 
\begin{quotation}
Do we then have to fall back on `no signalling faster than light' as the expression of the fundamental causal structure of contemporary theoretical physics? That is hard for me to accept. For one thing \textit{we have lost the idea that correlations can be explained}, or at least this idea awaits reformulation. [emphasis added]
\end{quotation}

Precisely what Bell meant here is perhaps not entirely clear. We take him to be saying that if we stop short of demanding of a causally well-behaved theory that it satisfy the factorisability condition (local causality) and impose only the no-signalling constraint (a recent suggestion of such an approach is \citet{ZukowskiandBrukner:2014} for example), then we are no longer in a position to explain, or to expect to explain, the quantum correlations. Such a view would compromise the motivation for the search for a locally causal stochastic completion of quantum mechanics. It would diminish the importance of the conclusion of post-1976 version of the Bell theorem, which is that locally causal, stochastic explanations of quantum correlations are, in the end, unavailable. 

How bad is that? Such a blunt question is partly motivated by Bell's own doubts: earlier in the 1990 paper he recommended that the factorizability condition `be viewed with the utmost suspicion'. Could it  be that his misgivings had to do with the fact that all the motivation for the condition provided in his papers is couched in abstract terms, with no guidance from concrete models? Bell himself stressed the importance of the role of models in checking our physical intuitions; in particular  he laid stress on study of the details of the de Broglie--Bohm theory in the context of theorising about the nature of deterministic hidden variable theories. The same kind of lesson was stressed by \citet{dickson:1998}. In this spirit, we think that the Everett picture of quantum mechanics can play an illuminating role in understanding the significance of factorizability. It is a model in which, remarkably, \textit{there is intrinsic unpredictability, but no strict indeterminism}, thus opening the door to a nuanced notion of randomness, and more imortantly, probability. The branching structure of the universe gives rise to a subtle reinterpretation of the notion of correlations between entangled systems. Most significantly, the Everett version of quantum mechanics provides a Lorentz covariant picture of the world in which factorizability fails but there is no action-at-a-distance.

\section{Locality in the Everett picture}\label{Everett}

Amongst those who have taken Everett's approach to quantum theory at all seriously as an option, it is a commonplace that---given an Everettian interpretation---quantum theory is (dynamically) local---there is no action-at-a-distance. Indeed this is often taken as one of the main selling-points of an Everettian approach (cf. \citet{lev:seop}, for example). Everett himself noted that his approach obviated the `fictitious paradox' of EPR (\citet[\S 5.3]{everett:1957}), though he didn't go into details, beyond noting the crucial role of the collapse of the wavefunction in the EPR argument---collapse which is of course absent in purely unitary Everettian quantum mechanics. \citet{page:1982} filled-in the details of how the absence of collapse in Everett circumvents the EPR dilemma. Other discussions of locality in Everett, covering also the extension to Bell's argument as well as EPR's, include amongst others \citet{vaidman:1994}, \citet{tipler:2000}, \citet{bacciagaluppi:2002}, and \citet{timpsonandbrown:2002}.\footnote{See also \citet[Chapter 8]{wallace:book}. Another important stream in the locality-in-Everett literature begins with \citet{dh} and their emphasis of locality in Everettian Heisenberg picture quantum mechanics. (See also \citep{rubin:2001,rubin:2002,rubin:2011,horsmanvedral:2007}.) However, Deutsch and Hayden's claims are best understood not as addressing the question of locality in the sense of no action-at-a-distance, but rather the distinct question of the \textit{separability of the states} of one's fundamental theory. See \citep{timpson:2005,wallacetimpson:2007,arntzenius:2012,wallacetimpson:2014}. Another example of an essentially Everettian treatment is the recent \citet{brassard:locality}.}

A number of interrelated factors are involved in Everett's theory being local.\footnote{Whilst we speak of \textit{Everett's theory} or of \textit{Everettian quantum mechanics} or of \textit{the Everrett interpretation}, it is of course true that Everett's original ideas have been developed in rather different directions by different authors. Nothing much in our present discussion of locality hangs on this, in our view, but for the record, we take the current canonical form of Everretianism to be that articulated in \citep{simon1,simon2,simonrelativism,simon3,saundersetal:2010,wallace:book}.} First, the point already noted, that by remaining within purely unitary quantum mechanics---thus eschewing collapse---one moves out of range of the EPR argument, even if one believes (\textit{contra} Bohr et al.) that the wavefunction should be understood realistically. Second, the point that the unitary dynamics itself derives from local Hamiltonians (in the relativistic context, moreover, Lorentz-covariant Hamiltonians) which include only point-like interactions. Third, the fact that the fundamental states of the theory are \textit{non-separable}, due to entanglement: it is not the case that the properties of a joint system are determined by the properties of the individual parts taken in isolation. And fourth, the crucial fact that following a measurement, it is generally not the case that there is one unique outcome of the measurement. It is this last, of course, which lies behind the `many worlds' conception of the theory: following a measurement interaction between a measuring apparatus and a system not in an eigenstate of the quantity being measured, the measuring apparatus will be left in a superposition of its indicator states. Unitary processes of decoherence lead to effective non-interaction between the different terms in this superposition, and lead to effective classicality for their future evolution over time. As further items interact with the apparatus---whether environmental (noise) degrees of freedom, or perhaps observers in the lab---they too become drawn linearly into the superposition, becoming part of the overall entangled state. An effective branching structure of macroscopically determinate goings-on emerges (at least at a suitably coarse-grained level) and we can reasonably call these emergent branches of effectively isolated, quasi-classical, macroscopically determinate goings-on, \textit{worlds}. Within particular branches in the structure will be found measuring apparatuses each giving definite readings for the outcome of experiments performed, and correlated with these can be found observers, each of whom will see their measuring device indicating a single particular result of measurement. But zooming-out (in a God's-eye view) from a particular branch will be seen all the other branches, each with a different result of measurement being recorded and observed, all coexisting equally; and all underpinned by (supervenient on) the deterministically, unitarily, evolving universal wavefunction.

Now why is this fourth factor---the non-uniqueness in the complete state of the world of the measurement outcomes for a given measurement---so significant when it comes to considering locality? Some commentators point out---plausibly---that it is simply an implicit premise in Bell's discussions that specified measurements, when performed, \textit{do} have unique outcomes: that the beables that obtain in a given region of spacetime where a measurement has taken place do fix a unique value for the outcome of that measurement (do fix, that is, a unique value for the quantity measured). When this implicit premise is given up, Bell's reasoning just doesn't apply, they note (\citet[p.310]{wallace:book}, \citet{blaylock}, \citet{maudlin:what})\footnote{In \citet{Wiseman:2014}, that measurements yield unique outcomes is stipulated in Axiom 1, p. 4; see also the first paragraph of section 7 therein.}. In that case, we can infer nothing as to the presence or absence of locality in a theory when it violates a Bell-inequality, or equivalently, when it violates factorisability.

This response is correct, so far as it goes, but i) it is perhaps less obvious how `unique outcomes' functions as an assumption in Bell's reasoning once he moved to consider stochastic theories and local causality as formulated from 1976 on, rather than the deterministic theories of 1964, and ii) one can hope to say a little more to illuminate the situation than just that this implicit premise fails. Indeed, regarding (ii) it is in our view rather helpful to go into some of the details of how in fact the Everett interpretation deals with the generation of correlations in EPR and Bell settings in a local manner, when this proves so difficult to do, or impossible to do, for other approaches. For convenience we will recapitulate here the analysis we gave in \citet{timpsonandbrown:2002} before going on to make some further points. We will return to address (i) briefly in Section~\ref{role of separability}. 

\subsection{EPR and Bell correlations in the Everettian setting}\label{everettiananalysis}

\sloppypar It is a straightforward matter to apply Everettian measurement theory in the context of EPR-Bell scenarios. (Very similar analyses can also be found elsewhere, for example in \citep{tipler:2000,tipler:2014}). We will discuss once again, by way of concrete example, the Bohm version of the EPR experiment, involving spin. We begin with two spin-1/2 systems (labelled 1 and 2) prepared in a singlet state. The relevant degrees of freedom (the `pointer variables') of measuring apparatuses $m_{A}, m_{B}$ in widely separated regions $A$ and $B$ of space we can model as two-state systems with basis states $\{\ket{\up}{A},\ket{\down}{A}\}$,\linebreak $\{\ket{\up}{B},\ket{\down}{B}\}$ respectively. Measurement by apparatus $A$ of the component of spin at an angle $\theta$ from the $z$-axis on system 1 would have the effect
\begin{equation}\label{measurement}
U({\theta}) \left\{ \begin{array}{ccc}
			\ket{\up_{\theta}}{1}\ket{\up_{\theta}}{A} & \mapsto & \ket{\up_{\theta}}{1}\ket{\up_{\theta}}{A} \\
			\ket{\down_{\theta}}{1}\ket{\up_{\theta}}{A} & \mapsto & \ket{\down_{\theta}}{1}\ket{\down_{\theta}}{A}      \end{array}
\right.,
\end{equation}
where $\ket{\up_{\theta}}{},\ket{\down_{\theta}}{}$ are eigenstates of spin in the rotated direction; similarly for measurement of system 2 by apparatus $B$.\footnote{Of course, for a measurement \textit{truly} to have taken place, the indicator states of the apparatus would also  need to have been irreversibly decohered by their environment, and themselves to be robust against decoherence. We can put these complications to one side for the purposes of our schematic model.}

\subsubsection{Case 1: Perfect correlations (EPR)}\label{perfect case}

Now consider the case in which our measuring apparatuses are perfectly aligned with one another, as in the EPR example. The initial state of the whole system is

\begin{equation}\label{parallel initial}
\ket{\up}{A}\tfrac{1}{\sqrt{2}}\bigl(\ket{\up}{1}\ket{\down}{2} - \ket{\down}{1}\ket{\up}{2}\bigr) \ket{\up}{B}.
\end{equation}
Note that the states of the measuring apparatuses factorise: at this stage, they are independent of the states of the spin-1/2 particles. For the measurements to be made, systems 1 and 2 are taken to regions $A$ and $B$ respectively and measurement interactions of the form (\ref{measurement}) occur (the time order of these interactions is immaterial); we finish with the state

\begin{equation}\label{parallel final}
\tfrac{1}{\sqrt{2}}\bigl( \ket{\up}{A}\ket{\up}{1}\ket{\down}{2}\ket{\down}{B}-\ket{\down}{A}\ket{\down}{1}\ket{\up}{2}\ket{\up}{B}\bigr).
\end{equation}

\textit{Now} definite states of the measuring device $m_{A}$ are correlated with definite spin states of particle 1, as are definite states of the measuring device $m_{B}$ with definite spin states of particle 2. But since all of the outcomes of the measurements are realised, there is no question of the obtaining of one definite value of spin at one side forcing the anti-correlated value of spin to be obtained at the far side. Both sets of values become realised, relative to different states of the apparatus (and relative to subsequent observers if they are introduced into the model)---as we put it before:
\begin{quotation}
\noindent There is, as it were, no dash to ensure agreement between the two sides to be a source of non-locality and potentially give rise to problems with Lorentz covariance.\citep{timpsonandbrown:2002}
\end{quotation}
However, it is important to note that following the measurements at $A$ and $B$, not only does each measured system have a definite spin state relative to the indicator state of the device that has measured it, but the systems and measuring apparatuses in each region (e.g. system 1 and apparatus $m_{A}$ in $A$) have definite spin and indicator states relative to definite spin and indicator states of the system and apparatus in the \textit{other} region (e.g. 2,$m_{B}$ in $B$). That is, following the two local measurements, from the point of view of the systems in one region, the states of the systems in the far region correspond to a definite, in fact perfectly anti-correlated, measurement outcome. This is in contrast to the general case of non-parallel spin measurements at $A$ and $B$, as we shall see in a moment.

In this parallel-settings, EPR, case we have a deterministic explanation of how the perfect (anti-)correlations come about. Given the initial state that was prepared, and given the measurements that were going to be performed, it was always going to be the case that a spin-up outome for system 1 would be correlated with a spin-down outcome for system 2 and \textit{vice versa}, once both sets of local measurements were completed. The perfect (anti-) correlations unfold deterministically from the initial entangled state given the local measurement interactions in regions $A$ and $B$ respectively. There is no puzzle here about how independent, intrinsically stochastic processes taking place in spacelike separated regions could nonetheless end up being correlated, since the evolution, and explanation of how the correlations come about, is purely deterministic: \textit{up} for system 1 was always going to be correlated with \textit{down} for system 2, as was \textit{down} for system 1 with \textit{up} for system 2, both cases (of course) being superposed in the final state.\footnote{Note that given that both possible outcomes for each spin measurements obtain simultaneously in the final state, the Bell--Suppes--Zanotti argument that any local theory which predicts perfect (anti-)correlations must collapse into the kind of deterministic theory Bell considered in his (1964), does not obtain. Everett is a deterministic theory, but it does not belong to the class of deterministic completions of quantum mechanics considered in \citet{bell:1964}, since it is \textit{also} a probabilistic (stochastic) theory---at the emergent level at which measurement outcomes are part of the story.}

\subsubsection{Case 2: Non-aligned measurements (Bell)}\label{Bell case}

As we have already noted, one of the plethora of good points in \citet{bell:1964} is of course that the parallel-settings case is in an important sense not very interesting. For this restricted class of measurement settings, Bell readily provided a local (deterministic) hidden variable model. Moreover, from a practical point of view, cases in which one has managed exactly to align one's Stern-Gerlach magnets with one's colleagues' way over yonder will form a set of measure zero. For both these reasons, then, it is important to consider what happens in the general case---as in the set-up required to derive Bell's inequalities---where the measurement angles are not aligned. 

In this case we may write the initial state as

\begin{equation}\label{non-parallel initial 1}
\ket{\up}{A}\tfrac{1}{\sqrt{2}}\bigl( \ket{\up}{1}\ket{\down}{2} - \ket{\down}{1}\ket{\up}{2}\bigr) \ket{\up_{\theta}}{B},
\end{equation}
where we have assumed a relative angle $\theta$ between the directions of measurement. If we write $\ket{\up}{}=\alpha \ket{\up_{\theta}}{}\! + \!\beta \ket{\down_{\theta}}{},\; \ket{\down}{}=\alpha^{\prime}\ket{\up_{\theta}}{}\! + \!\beta^{\prime}\ket{\down_{\theta}}{}$, we can express this joint state as:
\begin{equation}\label{non-parallel initial 2}
\ket{\up}{A}\tfrac{1}{\sqrt{2}}\Bigl[\, \ket{\up}{1} \bigl( \alpha \ket{\up_{\theta}}{2}\! +\! \beta \ket{\down_{\theta}}{2}\bigr) - \ket{\down}{1} \bigl( \alpha^{\prime}\ket{\up_{\theta}}{2}\! +\! \beta^{\prime}\ket{\down_{\theta}}{2} \bigr)\Bigr]\: \ket{\up_{\theta}}{B}.
\end{equation}
We can then see that following the measurements at $A$ and $B$ (the time order of these two measurements is again immaterial, of course), we will have

\begin{equation}\label{non-parallel middle}
\begin{split}
\frac{1}{\sqrt{2}} \Bigl[ \,\ket{\up}{A}\ket{\up}{1}\:\bigl(\alpha \ket{\up_{\theta}}{2}\ket{\up_{\theta}}{B}& + \beta \ket{\down_{\theta}}{2}\ket{\down_{\theta}}{B}\bigr)   \\
	 - \,&\ket{\down}{A}\ket{\down}{1}\,\bigl(\alpha^{\prime}\ket{\up_{\theta}}{2}\ket{\up_{\theta}}{B} + \beta^{\prime}\ket{\down_{\theta}}{2}\ket{\down_{\theta}}{B}\bigr)\Bigr].  
\end{split}
\end{equation} 
Here, relative to states representing a definite outcome of measurement in region $A$, there is no definite outcome in $B$; system 2 and apparatus $m_{B}$ are just entangled, with no definite spin and indicator states. Similarly, from the point of view of definite spin and indicator states of 2 and $m_{B}$ (a definite outcome in region $B$), there is no definite outcome of measurement in region $A$.

For non-parallel spin measurements, then, unlike the parallel case, there needs to be a third measurement (or measurement-like interaction), comparing (effectively) the outcomes from $A$ and $B$, in order to make definite spin and indicator states from one side definite relative to definite spin and indicator states from the other. Thus systems from $A$ and $B$ have to be brought (or come) together and a joint measurement be performed (or a  measurement-like interaction take place), leading to a state like:

\begin{multline}\label{non-parallel final}
\frac{1}{\sqrt{2}} \Bigl[\, \alpha \ket{\up}{A}\ket{\up}{1}\ket{\up_{\theta}}{2}\ket{\up_{\theta}}{B}\ket{\up\up}{C} + \beta \ket{\up}{A}\ket{\up}{1}\ket{\down_{\theta}}{2}\ket{\down_{\theta}}{B}\ket{\up\down}{C}  \\
 - \alpha^{\prime}\ket{\down}{A}\ket{\down}{1}\ket{\up_{\theta}}{2}\ket{\up_{\theta}}{B}\ket{\down\up}{C} - \beta^{\prime} \ket{\down}{A}\ket{\down}{1}\ket{\down_{\theta}}{2}\ket{\down_{\theta}}{B}\ket{\down\down}{C}\Bigr],  
\end{multline}     
where the states $\ket{\up\up}{C}$ etc. are the indicator states of the comparing apparatus. Following this third measurement-interaction, which can only take place in the overlap of the future light cones of the measurements at $A$ and $B$, a definite outcome for the spin measurement in one region finally obtains, relative to a definite outcome for the measurement in the other. That is, we can only think of the \textit{correlations} between measurement outcomes on the two sides of the experiment actually obtaining in the overlap of the future light-cones of the measurement events---they do not obtain before then and---\textit{a fortiori}--- they do not obtain instantaneously. On each side \textit{locally} there are definite measurement outcomes (superposed with one another) as soon as each local measurement is complete, but there is no correlation between the measurement outcomes on the two sides until later on, when a suitable entangling operation between systems from the two regions $A$ and $B$ can take place. 

It is important to note what this means for how probability statements derived from the Born rule for joint measurements on spacelike separated systems should be understood in this context. Given the initial entangled singlet state, we can make formal statements about what the probability distribution over joint measurement outcomes for spacelike separated measurements is. In general these joint-probabilities will be non-trivial and in fact Bell-inequality violating. But physically, there is nothing for such joint probabilities to be joint probabilities \textit{of} until one reaches the overlap region of the future light-cones of the measurement events, since it is only in this overlap region that measurement-outcomes for the individual measurements on either side can become definite with respect to one another, in general. Before then, the joint probabilities are only formal statements, regarding what one would expect to see, were one to compare the results of measurements on the two sides. This state of affairs does much to take the sting out of the randomness problem mentioned above in Section~\ref{unpacking}: the original separated measurement events are neither fundamentally random, nor are they correlated in the straightforward sense of classical physics.

\subsection{The role of non-separability}\label{role of separability}

Is there anything left to be explained? Arguably yes: it is the non-trivial fact that the correlations \textit{will} be found to obtain in the future, given a future comparison measurement.

So far we have primarly emphasised the role of the failure of uniqueness for measurement outcomes in permitting Everettian quantum mechanics to produce EPR and Bell correlations without any action-at-a-distance, and we have seen in some detail how in fact these correlations can be explained as arising following a local dynamics from the initial entangled state. Earlier, however, we noted that non-separability was also an important part of the explanation of how Everett could provide a local story for EPR and Bell correlations. We shall expand on this now, and in so-doing return to the point raised above that it may not be entirely obvious whether or how uniqueness of outcomes features as an assumption in Bell's reasoning from 1976 on, once the idea of local causality---or rather its formulation in terms of factorisability---was introduced.

One obvious point to make straightaway is that if the initial pure state of the two spins was not entangled, i.e., was separable, then there could not be \textit{any} correlation between the spin-measurements on the two sides of the experiment, let alone Bell-inequality violating ones. (This point holds in more general theories than quantum mechanics also.) But more deeply what is going on is that non-separability allows there to be facts about the relations between spatially separated systems which go above and beyond---are not determined by---the intrinsic (i.e., locally defined) properties of those systems individually. In particular, there can be facts about how things in spatial region $A$ will be correlated with (related to) things in spatial region $B$, without its being the case that how things are in $A$ and how things are in $B$ fix these relations.\footnote{Some early explorations of this idea in the context of Bell's theorem include \citet{howard:1985,howard:1989}, \citet{teller:1986,teller:1989}, \citet{french:1989}, \citet{healey:1991,healey:1994}, and \citet{mermin:ajp1998,mermin:1999,mermin:ithaca}.}

In the Bell-experiment with spins we have just discussed, if one only had failure of uniqueness of outcomes for the measurements on each side, in an otherwise separable theory, then there could obtain no non-trivial correlations between the outcomes of the measurements. Everettian quantum mechanics exploits both non-uniqueness of outcomes and non-separability in accounting for EPR and Bell correlations without action-at-a-distance. In fact, it is the particular way that non-separability features in the theory which entails non-uniqueness for the measurement outcomes.\footnote{We leave it as an open question whether or not in any non-separable theory which is dynamically local, but violates local causality, uniqueness of measurement outcomes fails.}

Now: the Everett interpretation shows that a theory can be local in the sense of satisfying no-action-at-a-distance, whilst failing to be locally causal: it violates the factorisability condition, or equivalently, the condition in terms of the probabilities attached to local beables in a given spacetime region being independent of goings-on in spacelike regions, once the state of the past light-cone of that region is sufficiently specified. What, then, went wrong with Bell's formulation of local causality, as an expression of locality? It is not perspicuous, at least to us, why the mere failure of uniqueness of measurement outcomes should make local causality---as Bell formulated it mathematically---inapposite as an expression of a principle of locality. So there is something more to be said, here. In our view what needs to be said is that, at root, \textit{it is moving to the context of non-separable theories} which makes Bell's mathematically formulated conditions fail properly to capture his intuitive notions of locality.\footnote{In making this claim we are thereby taking issue with some of the conclusions of the otherwise splendid discussion of \citet{henson:separability}. We will expand on our disagreement with Henson in full on another occasion.}

To substantiate this thought we need to return to Reichenbach's principle of the common cause. Recall that in order to move from Bell's informal statement of local causality---that the proximate causes of events should be nearby them, and causal chains leading up to these events should lie on or within their past light-cones---to its mathematical formulation in terms of factorisability or equivalently in terms of screening-off, something like Reichenbach's principle needs to be appealed to: statistical correlations between events must be explained either by direct causal links between them, or in terms of a Reichenbachian common-cause in the past. With such a principle in place we can connect mathematical statements in terms of probabilities to physical claims about causal links---and thereby to claims about action-at-a-distance. However, once we move to a situation where our theory can be non-separable, the common cause principle is unnatural and unmotivated.\footnote{\citet{seevinckuffink:bathwater} argue that we need not see Bell's mathematical conditions as resting on Reichenbach's principle, but on the notions of locality, causality and statistical sufficiency. In our view their analysis in terms of statistical sufficiency would also be uncompelling in the context of non-separable theories, for much the same reasons as Reichenbach's principle is. Again, we leave a detailed development of this claim to another forum.} This is for a very simple reason: in a non-separable theory there is a \textit{further} way in which correlations can be explained which Reichenbach's stipulations miss out: correlations between systems (e.g., the fact that certain correlations between measurement outcomes \textit{will} be found to obtain in the future)  can be explained directly by irreducible relational properties holding between the systems, relational properties themselves which can be further explained in dynamical terms as arising under local dynamics from a previous non-separable state for the total system. Which is precisely what happens in the Everettian context, for example.

In sum, we can see Bell's mathematical formulation(s) of the intuitive idea of local causality as instantiating Reichenbach's principle of the common cause\footnote{Modulo the historical subtleties noted earlier about whether in Reichenbach's own setting the principle should really be thought to apply at the micro-level at all.} as applied in the case of measurements on systems in spacelike separated regions, where we are assuming the standard (or naive) conception of what constraints relativity imposes on causal processes. It may be that Reichenbach's principle is plausible enough when one considers separable theories, but it is unacceptable when one considers non-separable theories. It is therefore, at the most straightforward level, simply because Reichenbach's principle is not apt for worlds which may be non-separable that Bell's formal statements of local causality go wrong, and that there can be theories such as Everett's which are not locally causal, but which are local, in the sense that they involve no action-at-a-distance.

We noted at the end of Section~\ref{unpacking} that Bell appeared to sound a note of despair at the end of Nouvelle Cuisine when contemplating the prospect that local causality might not turn out to be an adequate statement of locality in physics. In particular he worried that should his formulations of local causality not be apt and that we had instead to settle for some other statement of locality (such as no-signalling) which would allow the existence of Bell-inequality violating correlations whilst one's theory counted as fully local, then we would have `lost the idea that correlations can be explained' to quote him again. And this would indeed seem a worrying thing. But it seems to us, at least, that Bell need not have cause to despair in the circumstances which we have sketched. In our view, correlations which violate factorisability, whilst yet arising from a dynamically local theory, need not be condemned to be unexplainable: we just need to free ourselves from a Reichenbachian-common-cause straightjacket of what suitable explanation could be. Put another way, we can all actually fully \textit{agree} with the most basic sentiment which commentators draw from Reichenbach, namely that correlations should be explainable, whilst disagreeing with his specific formulation of what causal explanation (or maybe just \textit{explanation}) in terms of factors in the past must be like.\footnote{Accordingly, we commend the approach of \citet{cavalcantilal} who explicitly separate Reichenbach's principle into two components: First that correlations should be causally explicable either by direct interaction or common cause in the past; Second that explanation by common cause in the past takes the particular form that Reichenbach imposed. One can maintain the first idea whilst varying the second.} Specifically, it need not be the case that the factor in the past should be some classical random variable which screens-off the correlations. A perfectly acceptable, non-equivalent, alternative form of explanation of correlations in terms of factors in the past would seem to be in terms of the evolution of a later non-separable state from an earlier non-separable state. We saw an instance of this, of course, in the Everettian case earlier. The \textit{general} story as to how a non-separable theory can locally explain Bell-inequality violating correlations would be that the correlations are entailed by some suitable non-separable joint state. And if one has a detailed story of the contents of one's local and non-separable theory (i.e., a detailed specification of its kinematics and dynamics---including measurement theory) then one will have a perfectly good explanation of how this comes about. One's explanation will be in terms of how that particular non-separable joint-state evolved out of some previous (generally) non-separable state.\footnote{Thus we offer a picture in terms of the fundamental states of a theory and their dynamical---law-governed---evolution. Is this a \textit{causal} form of explanation? If it is thought not, to our minds it is not clear that that matters. It is certainly \textit{physical explanation} by one gold standard.}                 

\subsection{Maudlin's challenge}   

We have sought to explain how the Everett interpretation provides one concrete example illustrating that Bell's mathematical, probabilistic, formulation of local causality does, as he feared it might, fail adequately to capture the notion of locality. An important challenge to a conclusion of this kind is presented in characteristically trenchant and pithy form by \citet{maudlin:what} however, to which we must now turn.

Maudlin argues that:
\begin{quote}
Because the many worlds interpretation fails to make The Predictions [a subset of the predictions of standard quantum theory], Bell's theorem has nothing to say about it. ... And reciprocally, the existence of the many worlds interpretation can in principle shed no light on Bell's reasoning because it falls outside the scope of his concerns. \citep[\S III]{maudlin:what}
\end{quote}
We will consider these claims in turn.

The first thing Maudlin includes in `The Predictions' is the claim that measurements (of spin, for the EPR--Bohm experiment we have been considering) have unique outcomes, where `Born's rule provides the means of calculating the probability of each of the...outcomes.' (\textit{ibid.}). The further elements are the prediction of perfect EPR correlations for parallel measurements, and the prediction of Bell-inequality violating correlations for certain specific choices of non-aligned measurements. It is because he sees uniqueness of measurement outcomes as necessary conditions for the further predictions of EPR and Bell correlations to have any content that Maudlin believes the Everett interpretation gets into trouble:    

\begin{quote}
If every experiment carried out on particle 1 yields both [possible outcomes] ... and every experiment  carried out on particle 2 yields both [possible outcomes], what can it mean to say that the outcomes on the two sides are always correlated or always anticorrelated [as in the EPR--Bohm scenario] or agree only [some percentage] ... of the time [as in the EPR--Bell scenario]? For such claims to have any content, particular results on one side must be associated with particular results on the other so that the terms `agree' and `disagree' make sense. \citep[\S III]{maudlin:what}
\end{quote}

The first thing to note is that from the point of view which we have been entertaining, it is simply question-begging to include `uniqueness of measurement outcomes' amongst the predictions of \textit{standard quantum theory}. For what, after all, even \textit{is} standard quantum theory? Arguably there is no such thing---there is a standard quantum \textit{algorithm}, which experimentalists know how to apply to get a good fit to experiment, and where in particular they find that the Born rule gives an excellent fit to the results they see. But anything beyond this is up for grabs, and may be understood differently in different interpretations (or different theories) offered to underpin the success of the standard quantum algorithm. In particular, from an Everettian point of view, the \textit{observation} of unique outcomes of measurements doesn't at all entail \textit{uniqueness of the outcomes in the complete state of the world}. For the Everettian, the success of the standard quantum algorithm can be guaranteed without requiring the latter form of uniqueness.   

But even if uniqueness of measurement outcomes can't be thought to be part of the \textit{predictions} of standard quantum theory (as opposed to being part of the presuppositions in various standard approaches to the theory) it might even so---as Maudlin alleges---be a necessary condition for making sense of the prediction of correlations, for the reasons Maudlin states above. We disagree, however.

First return to the Everettian treatment of perfect EPR correlations as we described above (Section~\ref{perfect case}). Here we saw how the standard, local, unitary dynamics would lead deterministically to the case in which \textit{up} on one side of the experiment was correlated with \textit{down} on the other, and \textit{vice versa}. Relative to definite results on one side of the experiment, there are definite results on the other. It is quite clear what it means for the measurement results always to be correlated in this case, notwithstanding the fact that both options for the correlated outcomes are superposed together in the overall final entangled state. Note for future reference that in this case we do not have to appeal to the Born rule to understand the prediction that the results on the two sides of the experiment will be perfectly correlated with one another.

Second, we return to our treatment of the non-aligned, Bell, case (Section~\ref{Bell case}). Here one will need to appeal to the Born rule in order to predict that there are correlations, but importantly, as we described, it is only in the overlap of the future light-cones of the measurement events, given a suitable comparison interaction between systems from the two sides, that measurement outcomes from one side will become definite relative to measurement outcomes from the other. Thus what is required to make sense of the prediction of Bell correlations is that a measurement in the overlap of the future light-cones of the initial measurement events, one which compares locally, at a point, records of the outcomes from the two regions $A$ and $B$ where the initial measurements took place, should give a suitable probability distribution for what the results of that local comparison will be. In the non-aligned case, that `particular results on one side [are] associated with particular results on the other' comes about subsequent to the intial measurements being made, and is brought about by their being a suitable later comparison interaction. It is after this further interaction that `the terms ``agree'' and ``disagree'' make sense' as applied to the results of the earlier measurements.

It is possible that one might remain unsatisfied by this. After all, one might say that---at least for the Bell-correlations case, if not the EPR---we have had to appeal to the Born rule, and perhaps the real thrust of the worry coming from non-uniqueness of measurement outcomes is that sense can't be made of how the Born rule could apply to govern the probabilities of outcomes of measurements, if all the outcomes occur. This, of course, is a long-standing and respectable objection to the Everett interpretation, but notice \textit{that there is nothing specific about EPR or Bell correlations in this}. Given our analysis, the case of alleged difficulty in predicting Bell correlations has been reduced to the case of making sense of probability in Everett for local measurements. And in our view, this amounts to reduction to---plausibly---a previously solved case (see particularly the work of Saunders, Deutsch, Wallace and Greaves).\footnote{\citet{simon3}, \citet{deutsch:1999}, \citet{wallace:2003,wallace:2007,wallace:howtoprove}, \citet{greaves:2004} and \citet[Part {II}]{wallace:book}.} Even if one remains agnostic about whether or not Everett does give an adequate account of probability for local measurements, then our point is simply that \textit{in so far} as Everett is a player at all as a viable interpretation of quantum theory, it provides a concrete counterexample to Bell's probabilistic formulation of local causality. (In our view, of course, Everett \textit{is} a very significant player.)\footnote{There is a final worry which might be motivating Maudlin, in that he may find it obscure how the Everettian story about the emergence of determinate--but nonetheless superposed---measurement outcomes and experiences of the everyday world is supposed to work. But again this is just a general objection to Everett, and one on which there has been a great deal of persuasive work, see again \citet{saundersetal:2010} and \citet{wallace:book} for the state of the art.}

In effect we have already stated our response to Maudlin's reciprocal claim that `the existence of the many worlds interpretation can shed no light on Bell's reasoning because it falls outside the scope of his concerns'. On the contrary. Once it is recognised that the uniqueness of measurement outcomes is question-begging if assumed to be a requirement on there being well-formed theories which make probabilistic predictions for the results of measurements in various spacetime regions, Everett plainly falls within the scope of Bell's post-1976 reasoning and is entirely germane to the adequacy of his probabilistic formulation of local causality. In 1986 Bell noted:
\begin{quote}
The `many world interpretation' seems to me an extravagant[...]hypothesis[...]And yet...It may have something distinctive to say in connection with the `Einstein Podolsky Rosen puzzle' \citep{bell:six}
\end{quote}
We think he was dead right on this last point.

\section{Conclusions}

We have seen that there was a very significant shift in Bell's notion of nonlocality between 1964 and 1976, a shift that occurred in parallel with his thinking moving away from focussing on completions of quantum mechanics in the spirit of EPR. Crucially, violating the 1964 locality condition gives rise to action-at-a-distance, whereas violating local causality of 1976 need not. We have noted, moreover, that as Bell's thinking developed, he came more and more to recognise that there need be no straightforward conflict between violation of either of his locality conditions and the demands of relativity.

In our discussion of locality in the Everett interpretation we have sought to provide a constructive example illustrating precisely how a theory can be dynamically local, whilst violating local causality; and we emphasised the interconnected roles of the failure of uniqueness of measurement outcomes and of non-separability in achieving this. We think that Bell was right to have had doubts in 1990 regarding whether he had managed, in his mathematical, probabilistic, statements of local causality adequately to capture the concept of locality. But we have suggested that even if local causality is rejected as the expression of locality, it need not follow that one is doomed to having to put up with unexplained correlations, as Bell feared one might be. For as we have explained, non-separable theories allow additional ways in which correlations can be causally explained without action-at-a-distance.

\section*{Acknowledgements}

We would like to thank the editor, Shan Gao, for the invitation to contribute to this auspicious volume celebrating the 50th anniversary of Bell's seminal 1964 paper. We would also like to thank Ray Lal, Owen Maroney, Rob Spekkens and David Wallace for helpful discussion. CGT's work on this paper was partially supported by a grant from the Templeton World Charity Foundation.

%\bibliographystyle{cambridgeauthordate}
%\bibliography{bell,bell2}

\end{document}